\documentstyle[twoside,psfig]{article}

\catcode`\@=11
\long\def\@makefntext#1{
\protect\noindent \hbox to 3.2pt {\hskip-.9pt  
$^{{\eightrm\@thefnmark}}$\hfil}#1\hfill}		

\def\@makefnmark{\hbox to 0pt{$^{\@thefnmark}$\hss}}	
	
\def\ps@myheadings{\let\@mkboth\@gobbletwo
\def\@oddhead{\hbox{}
\rightmark\hfil\eightrm\thepage}   
\def\@oddfoot{}\def\@evenhead{\eightrm\thepage\hfil
\leftmark\hbox{}}\def\@evenfoot{}
\def\sectionmark##1{}\def\subsectionmark##1{}}

\def\be{\begin{equation}}
\def\ee{\end{equation}}
\oddsidemargin=\evensidemargin
\newcounter{sectionc}\newcounter{subsectionc}\newcounter{subsubsectionc}
\renewcommand{\section}[1] {\vspace{12pt}\addtocounter{sectionc}{1} 
\setcounter{subsectionc}{0}\setcounter{subsubsectionc}{0}\noindent 
	{\tenbf\thesectionc. #1}\par\vspace{5pt}}
\renewcommand{\subsection}[1] {\vspace{12pt}\addtocounter{subsectionc}{1} 
	\setcounter{subsubsectionc}{0}\noindent 
	{\bf\thesectionc.\thesubsectionc. {\kern1pt \bfit #1}}\par\vspace{5pt}}
\renewcommand{\subsubsection}[1] {\vspace{12pt}\addtocounter{subsubsectionc}{1}
	\noindent{\tenrm\thesectionc.\thesubsectionc.\thesubsubsectionc.
	{\kern1pt \tenit #1}}\par\vspace{5pt}}
\newcommand{\nonumsection}[1] {\vspace{12pt}\noindent{\tenbf #1}
	\par\vspace{5pt}}

\newcounter{appendixc}
\newcounter{subappendixc}[appendixc]
\newcounter{subsubappendixc}[subappendixc]
\renewcommand{\thesubappendixc}{\Alph{appendixc}.\arabic{subappendixc}}
\renewcommand{\thesubsubappendixc}
	{\Alph{appendixc}.\arabic{subappendixc}.\arabic{subsubappendixc}}

\renewcommand{\appendix}[1] {\vspace{12pt}
        \refstepcounter{appendixc}
        \setcounter{figure}{0}
        \setcounter{table}{0}
        \setcounter{lemma}{0}
        \setcounter{theorem}{0}
        \setcounter{corollary}{0}
        \setcounter{definition}{0}
        \setcounter{equation}{0}
        \renewcommand{\thefigure}{\Alph{appendixc}.\arabic{figure}}
        \renewcommand{\thetable}{\Alph{appendixc}.\arabic{table}}
        \renewcommand{\theappendixc}{\Alph{appendixc}}
        \renewcommand{\thelemma}{\Alph{appendixc}.\arabic{lemma}}
        \renewcommand{\thetheorem}{\Alph{appendixc}.\arabic{theorem}}
        \renewcommand{\thedefinition}{\Alph{appendixc}.\arabic{definition}}
        \renewcommand{\thecorollary}{\Alph{appendixc}.\arabic{corollary}}
        \renewcommand{\theequation}{\Alph{appendixc}.\arabic{equation}}
        \noindent{\tenbf Appendix \theappendixc #1}\par\vspace{5pt}}
\newcommand{\subappendix}[1] {\vspace{12pt}
        \refstepcounter{subappendixc}
        \noindent{\bf Appendix \thesubappendixc. {\kern1pt \bfit #1}}
	\par\vspace{5pt}}
\newcommand{\subsubappendix}[1] {\vspace{12pt}
        \refstepcounter{subsubappendixc}
        \noindent{\rm Appendix \thesubsubappendixc. {\kern1pt \tenit #1}}
	\par\vspace{5pt}}

\newcommand{\textlineskip}{\baselineskip=13pt}
\newcommand{\smalllineskip}{\baselineskip=10pt}

\def\eightcirc{
\begin{picture}(0,0)
\put(4.4,1.8){\circle{6.5}}
\end{picture}}
\def\eightcopyright{\eightcirc\kern2.7pt\hbox{\eightrm c}}

\def\abstracts#1#2#3{{
	\centering{\begin{minipage}{4.5in}\footnotesize\baselineskip=10pt
	\parindent=0pt #1\par 
	\parindent=15pt #2\par
	\parindent=15pt #3
	\end{minipage}}\par}} 

\newcommand{\bibit}{\nineit}
\newcommand{\bibbf}{\ninebf}
\renewenvironment{thebibliography}[1]
	{\frenchspacing
	 \ninerm\baselineskip=11pt
	 \begin{list}{\arabic{enumi}.}
        {\usecounter{enumi}\setlength{\parsep}{0pt}     
	 \setlength{\leftmargin 12.7pt}{\rightmargin 0pt} 
         \setlength{\itemsep}{0pt} \settowidth
	{\labelwidth}{#1.}\sloppy}}{\end{list}}

\newcounter{itemlistc}
\newcounter{romanlistc}
\newcounter{alphlistc}
\newcounter{arabiclistc}

\newcommand{\fcaption}[1]{
        \refstepcounter{figure}
        \setbox\@tempboxa = \hbox{\footnotesize Fig.~\thefigure. #1}
        \ifdim \wd\@tempboxa > 5in
           {\begin{center}
        \parbox{5in}{\footnotesize\smalllineskip Fig.~\thefigure. #1}
            \end{center}}
        \else
             {\begin{center}
             {\footnotesize Fig.~\thefigure. #1}
              \end{center}}
        \fi}

\newcommand{\tcaption}[1]{
        \refstepcounter{table}
        \setbox\@tempboxa = \hbox{\footnotesize Table~\thetable. #1}
        \ifdim \wd\@tempboxa > 5in
           {\begin{center}
        \parbox{5in}{\footnotesize\smalllineskip Table~\thetable. #1}
            \end{center}}
        \else
             {\begin{center}
             {\footnotesize Table~\thetable. #1}
              \end{center}}
        \fi}

\def\@citex[#1]#2{\if@filesw\immediate\write\@auxout
	{\string\citation{#2}}\fi
\def\@citea{}\@cite{\@for\@citeb:=#2\do
	{\@citea\def\@citea{,}\@ifundefined
	{b@\@citeb}{{\bf ?}\@warning
	{Citation `\@citeb' on page \thepage \space undefined}}
	{\csname b@\@citeb\endcsname}}}{#1}}

\newif\if@cghi
\def\cite{\@cghitrue\@ifnextchar [{\@tempswatrue
	\@citex}{\@tempswafalse\@citex[]}}
\def\citelow{\@cghifalse\@ifnextchar [{\@tempswatrue
	\@citex}{\@tempswafalse\@citex[]}}
\def\@cite#1#2{{$\null^{#1}$\if@tempswa\typeout
	{IJCGA warning: optional citation argument 
	ignored: `#2'} \fi}}

\def\pmb#1{\setbox0=\hbox{#1}
	\kern-.025em\copy0\kern-\wd0
	\kern.05em\copy0\kern-\wd0
	\kern-.025em\raise.0433em\box0}

\def\fnt#1#2{\footnotetext{\kern-.3em
	{$^{\mbox{\scriptsize #1}}$}{#2}}}

\def\fpage#1{\begingroup
\voffset=.3in
\thispagestyle{empty}\begin{table}[b]\centerline{\footnotesize #1}
	\end{table}\endgroup}

\def\runninghead#1#2{\pagestyle{myheadings}
\markboth{{\protect\footnotesize\it{\quad #1}}\hfill}
{\hfill{\protect\footnotesize\it{#2\quad}}}}
\font\tenrm=cmr10
\font\tenit=cmti10 
\font\tenbf=cmbx10
\font\bfit=cmbxti10 at 10pt
\font\ninerm=cmr9
\font\nineit=cmti9
\font\ninebf=cmbx9
\font\eightrm=cmr8

\def\qed{\hbox{${\vcenter{\vbox{			
   \hrule height 0.4pt\hbox{\vrule width 0.4pt height 6pt
   \kern5pt\vrule width 0.4pt}\hrule height 0.4pt}}}$}}

\begin{document}
\setlength{\textheight}{7.7truein}  

\runninghead{T R Govindarajan}{Information from quantum blackhole physics}

\normalsize\textlineskip
\thispagestyle{empty}
\setcounter{page}{1}


\vspace*{0.88truein}
\fpage{1}
\rightline{\small IMSc/03/08/27}

\centerline{\bf Information from quantum blackhole physics
\footnote
{Talk presented at the conference
\lq\lq Space-time and Fundamental Interactions: Quantum Aspects''
in honour of A.P.~Balachandran's 65th birthday, Vietri sul Mare, Salerno, Italy
26th-31st May, 2003}}
\vspace*{0.37truein}
\centerline{\footnotesize T R Govindarajan}
\baselineskip=12pt
\centerline{\footnotesize\it The Institute of Mathematical Sciences, 
Tharamani}
\baselineskip=10pt
\centerline{\footnotesize\it Chennai 600 113, INDIA}
\vspace*{10pt}


\vspace*{0.21truein}
\abstracts{
The study of BTZ blackhole physics and the cosmological horizon of 3D
de Sitter spaces are carried out in unified way using the
connections to the Chern Simons theory on three manifolds with boundary.
The relations to CFT on the boundary is exploited to construct
exact partition functions and obtain logarithmic corrections to Bekenstein
formula in the asymptotic regime. Comments are made on the
dS/CFT correspondence frising from these studies.}{}{}


\vspace*{1pt}\textlineskip	
\section{Introduction}
For the past 10 years there has been tremendous progress in the
understanding of quantum blackhole physics. Progress is seen
at the attempts to understand the microstates which account for
the entropy of blackhole which naturally emerges when we study the
the quantum nature of the blackholes\cite{stringbh}. The classical studies of
Hawking and Bekenstein led to the well known formula\cite{bekhawk}:
 
\be
T~=~\frac{\kappa}{2\pi}~\qquad ;~\qquad S~=~\frac{Area}{4\pi}
\ee
 
All attempts to construct quantum theory of gravity string theory or canonical formulation
or semiclassical methods, brick wall method etc uses this crucial
data as the testing ground for the program.
Similar situation arises in the study of quantum gravity in de Sitter spaces
also even though the questions in this case are more complicated and
far from good understanding. String theoretic considerations led to
interesting link between AdS gravity and Conformal field theory on
the boundary\cite{adscft}. Herealso blackhole physics plays a good testing ground.
On the other hand in the case of de Sitter gravity, String theory is yet to come
out with suitable background to make any progress and relate to
CFT on the boundary.
 
Observational evidence suggests the positive cosmological constant
which has created interest in the gravitational dynamics of de Sitter space.
De Sitter space which is maximally symmetric $n$ dimensional space-time with positive
cosmological constant has a cosmological horizon and regions of
space-time which are not accessible to the observer. And the thermodynamics
of such a horizon is similar to that of blackhole horizon.
The difficulties with String theory in accommodating de Sitter space\cite{bousso}
and other quantisation procedures like loop gravity\cite{abhay}
in getting semiclassical descriptions forces one to look for
new arena where some non perturbative quantum gravity
answers  can be provided. 3D gravity falls in such a category
because of the connections to the Chern Simons gauge theory.
The bonus is the known connection to Wess-Zumino model on the
boundary will provide the required correspondences\cite{witten}.
 
Here we focus on 3D gravity with cosmological constant
with both positive and negative signatures corresponding
to Anti deSitter and de Sitter spaces. In the first case
we consider the BTZ blackhole and cosmological horizon in the second case.
We treat the horizons as boundaries.
 
In the second section we point out that the self adjoint extension
required for the definitions of Hamiltonians lead to states
localised near the horizon which serve as the appropriate
blackhole states. Then we consider BTZ BH and deSitter gravity
in a unfied way pointing out similarities and differences
and set up the details of calculations for the partition
function using CS gauge theory. In the fourth section
we do explicite computation and obtain the leading
Bekenstein Hawking contribution for the
entropy as well as subleading logarithmic corrections.
In the last section we point out the results in connections
with AdS/CFT or dS/CFT correspondences. We also remark
about issues in connection with Quasi normal modes and CFT
etc. And we conclude with discussions on our work in relation to other
contributions\cite{btzlog}.

\vspace*{1pt}\textlineskip  
\section{Self adjoint extensions and horizon states}
We study the time-independent modes of a massless scalar field in
various black hole backgrounds.
 
We find that for the non-extremal
black holes, in the near-horizon limit is described by the
Hamiltonian\cite{horizon}
\begin{equation}
\left(-\frac{d^2}{dx^2}  - \frac{1}{4x^2} \right) \chi = 0,
\label{conformal}
\end{equation}
where $x=(r - r_{+})$ is the near-horizon coordinate. $r_{+}$ is the
horizon. For the extremal Reissner-Nordstrom solution, however,
we get  near the horizon
\begin{equation}
 - \frac{d^2 \chi}{dx^2} = 0.
\label{spham}
\end{equation}
 
Another situation where we see the same equation is the near horizon
geometry of the one-dimensional black hole discovered by Witten
\cite{witten2d}. The same behaviour is exhibited
in the BTZ black
hole in $(2+1)-D$ gravity \cite{btz}.
In all these cases barring the extremal black holes,
(\ref{conformal}) is the near-horizon equation for the zero-mode
solution.
 
This Hamiltonian $H$ is a special case of a more general Hamiltonian
studied extensively in the literature\cite{horizon}. It is defined on a domain
$L^2 {\bf R}^+ , dx]$ and is of the form
\begin{equation}
H_\alpha = -\frac{d}{dx^2} + \frac{\alpha}{x^2}.
\label{halpha}
\end{equation}
Classically, the system described by this Hamiltonian is scale
invariant ($\alpha$ is a dimensionless constant). However, the quantum
analysis of this operator is much more subtle.
Proper analysis points out the existence of states localised
near the boundary (in this case Horizon) and scale invariance is broken.
These are the states to  be identified as blackhole states.

\vspace*{1pt}\textlineskip 
\section{3D gravity, BTZ blackhole and de Sitter space}

The gravity action $I_{grav}$
written in a first-order formalism (using triads $e$ and spin connection
$\omega$) is the difference of two Chern-Simons actions.
\begin{eqnarray}
I_{\hbox{\scriptsize grav}}
 ~ = ~I_{\hbox{\scriptsize CS}}[A]~ - ~I_{\hbox{\scriptsize CS}}[\bar A] ,
\label{b9}
\end{eqnarray}
where
\begin{eqnarray}
A ~=~ \left(\omega^a + \frac{i}{l}~ e^a\right) T_a , \qquad
\bar A ~=~ \left(\omega^a - \frac{i}{l}~ e^a\right) T_a
\label{b8}
\end{eqnarray}
are  $\hbox{SL}(2,{\bf C})$ or $ SU(2)\otimes SU(2)$
gauge fields (with $T_a=-i\sigma_a/2$).
Here, the cosmological constant $\Lambda ~=~ \pm (1/{l^2})$.
The Chern-Simons action $I_{\hbox{\scriptsize CS}}[A]$ is
\begin{eqnarray}
I_{\hbox{\scriptsize CS}} ~=~ {k\over4\pi}\int_M
  \hbox{Tr}\left( A\wedge dA + {2\over3}A\wedge A\wedge A \right)
\label{cs}
\end{eqnarray}
and the Chern-Simons coupling constant is
$k = l/4G$. 
The Chern-Simons coupling constant $k = - l/4G$; Lorentzian gravity is
obtained from the Euclidean theory by a continuation $G \rightarrow -G$.

Now, for a manifold with boundary, the Chern-Simons field theory is
described by a Wess-Zumino conformal field theory on the boundary.   
We are interested in computing the entropy of the Euclidean BTZ black hole/
de Sitter space if cosmological constant is positive.
The Euclidean continuation of the BTZ black hole as well as the de Sitter space 
has the topology of a solid torus \cite{btzlog,suneeta}.
The metric for the Euclidean BTZ black hole in the usual Schwarzschild-like coordinates is
\begin{eqnarray}
ds^2~=~N^2~d\tau^2 + N^{-2}~dr^2 + r^2~(d\phi + N^{\phi}d\tau)^2
\label{metdef4}
\end{eqnarray}
\noindent where $\tau$ here is the Euclidean time coordinate and
\begin{eqnarray}
N~=~\left(-M~+~\frac{r^2}{l^2}~-~\frac{J^2}{4r^2}\right)^{\frac{1}{2}}~,
~~~~~N_{\phi}~=~-\frac{J}{2r^2}
\end{eqnarray}                                                                 
The black hole metric is just the metric for hyperbolic three-space ${\cal H}_3$
\begin{eqnarray}
ds^2~ =~ \frac{l^2}{z^2} ~(dx^2+dy^2+dz^2),\quad \quad z>0 ,
\label{a4}
\end{eqnarray} 
Global $(2+1)-d$ de Sitter spacetime  is described by the metric
\begin{eqnarray}
ds^2 ~=~ - l^2 d\tau^{2} + l^2 \cosh^{2} \tau d\Omega^2
\label{met}
\end{eqnarray}
Equal time sections of this
metric are two-spheres, and there are no globally timelike Killing
vectors.
However, there does exist a timelike Killing vector in certain patches of
this spacetime. Figure 1 shows the Penrose diagram of
global de Sitter space with these patches - II and IV. These regions
are causally disconnected and the timelike Killing vector flows in opposite
directions in these two patches. Each of these patches is bounded by the
cosmological horizon, and described by the metric
\begin{eqnarray}
ds^2~=~ - N^2~dt^2 + N^{-2}~dr^2 + r^2~d\phi^{2}
\end{eqnarray}
where
\begin{eqnarray}
N^{2}~=~(1~-~\frac{r^2}{l^2}),
\label{lapse}
\end{eqnarray}
and $0\leq r \leq l$. $\phi$ is an angular coordinate with
period $2\pi$. The cosmological horizon in these coordinates is
at $r=l$. Constant $t$ surfaces are discs $D_{2}$,
and the topology of the patch is $D_{2} \otimes R$.
\begin{figure}[htbp] 
\vspace*{13pt}
\centerline{\psfig{file=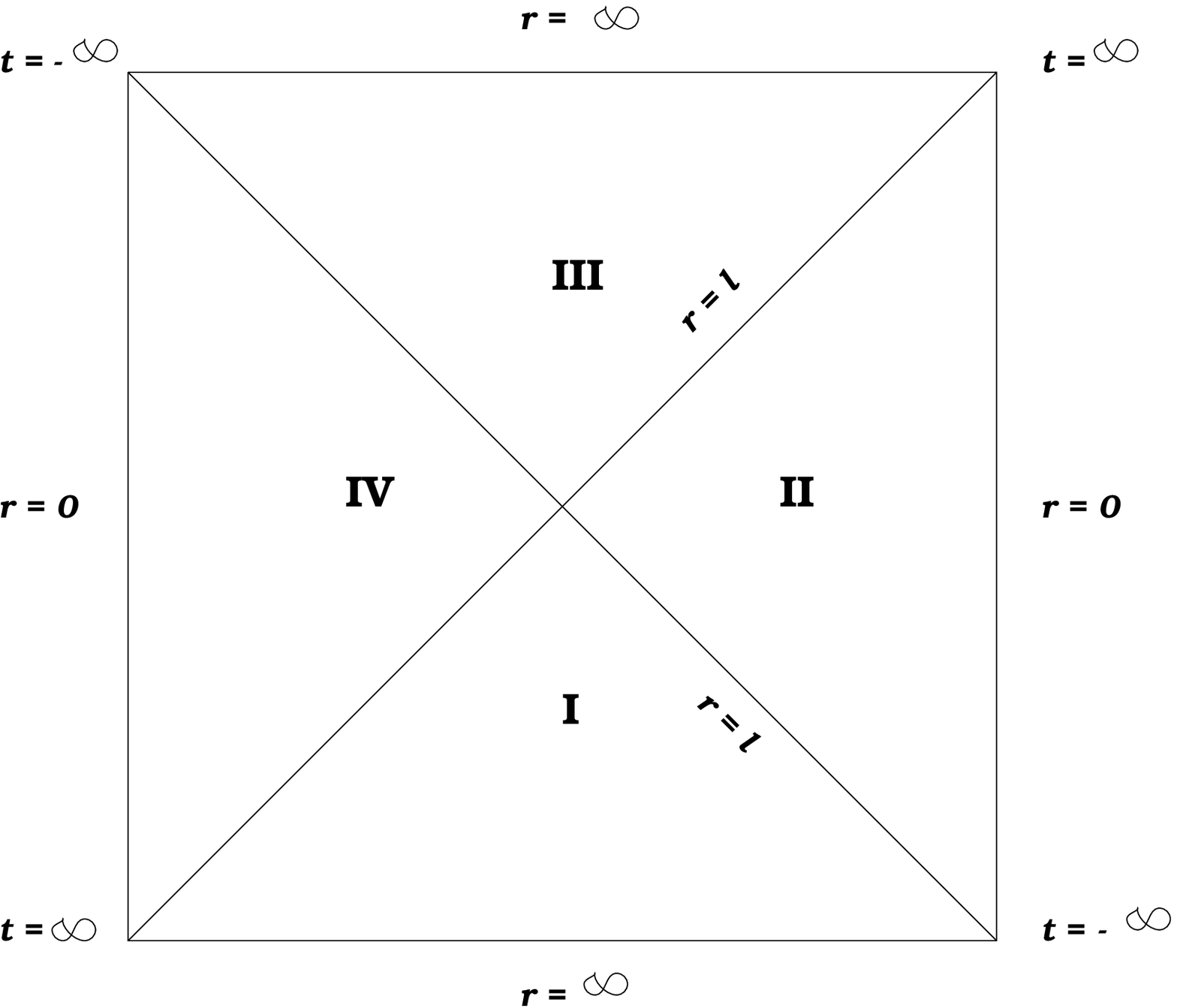}} 
\vspace*{13pt}
\fcaption{Penrose Diagram }
\end{figure} 

However, making the time periodic does convert it from $D_{2} \otimes R$ to a
solid torus $D_{2} \otimes S^{1}$ for each static patch II and IV.
The interpretation of the Euclidean static patch metric
covering the three-sphere completely (with the $r$ coordinate range being
covered twice\cite{suneeta} - {\em once} for each patch II and IV)
is geometrically the
gluing of the two solid tori (corresponding to the two patches II and IV)
in such a way that the resultant
manifold is closed; and a three-sphere.
This can be done easily by gluing the two solid tori with oppositely oriented
boundaries after performing a modular transformation on the boundary of one
of them.                                                       

\vspace*{1pt}\textlineskip 
\section{Computation of Partition function}

The connection can be written as \cite{labas}:
\begin{eqnarray}
A ~=~ \left(\frac{-i \pi \tilde u}{ \tau_2}~ d\bar z + \frac{i \pi u}{ \tau_2}
~dz\right) T_3
\label{adefn}
\end{eqnarray}
 
\noindent where $u$ and $\tilde u$ are canonically conjugate fields and obey
the canonical commutation relation:
\begin{eqnarray}
  [\tilde u, u]~ =~ \frac{2 \tau_2}{\pi (k+2)}
\label{ccr}
\end{eqnarray}
They can be related to the black hole parameters by computing the holonomies of
$A$ around the contractible and non-contractible
cycles of the solid torus.
Then the trace of the holonomies around the contractible cycle $A$ and
non-contractible cycle $B$ are:
\begin{equation}
Tr(H_A)~ =~ 2 \cosh (i \Theta), ~~~~~Tr(H_B)~ =~ 2 \cosh \left(\frac{2\pi}{l} (r_+
+ i |r_-|)\right)\label{udefn}
\end{equation}
and for de Sitter space
\begin{eqnarray}
Tr(H_a)~ =~ 2 \cos (\Theta), ~~~~~Tr(H_b)~ =~
2 \cos \left(\frac{\beta}{l}
\right)
\label{trh}
\end{eqnarray}

Now, we  write the Chern-Simons path integral on a solid torus with a
boundary modular parameter $\tau$. For
a fixed boundary value of the connection, i.e. a fixed value of $u$, 
this path integral is formally
equivalent to a state $\psi_{0}(u, \tau)$ with no Wilson lines in the solid torus.
The states corresponding  to having closed Wilson lines (along the
non-contractible cycle) carrying spin $j/2$
($j \le k$) representations in the solid torus
are given by \cite{labas}, \cite{elit} :
\begin{eqnarray}
\psi_{j}(u, \tau)~ =~ \exp\left\{ {\pi k\over4\tau_2}\,u^2 \right\}
~\chi_{j}(u, \tau) ,
\label{c3}
\end{eqnarray}
where $\chi_{j} $ are the Weyl-Kac characters for affine $\hbox{SU}(2)$.
The Weyl-Kac characters can be expressed in terms of the well-known
Theta functions as
\begin{eqnarray}
\chi_{j}(u, \tau)~ =~ \frac{\Theta_{j +1}^{(k+2)}(u, \tau, 0)~-~\Theta_{-j -1}^{(k+2)}(u, \tau, 0)}
{\Theta_{1}^{2}(u, \tau, 0)~-~\Theta_{-1}^{2}(u, \tau, 0)}
\label{c4}
\end{eqnarray}
where Theta functions are given by:
\begin{eqnarray}
\Theta_{\mu}^{k}(u, \tau, z) ~=~
\exp (-2 \pi i k z)~ \sum_{n \in \cal Z} \exp 2 \pi i k \left[(n + \frac{\mu}
{2 k})^2 \tau ~+~(n + \frac{\mu}{2 k}) u \right]
\label{theta}
\end{eqnarray}
For both  BTZ black hole and the
de Sitter partition function are constructed from the boundary state
$\psi_{0}(u, \tau)$. The construction is motivated by the
following observations :
 
(a) In the Chern-Simons functional integral over a solid torus, we shall
integrate over all gauge connections with fixed holonomy $H_b$ around
the non-contractible cycle. This corresponds to the partition function
with fixed period $\beta$ of the Euclidean time, that is, fixed inverse
temperature. This in
turn means we are dealing with the canonical ensemble.
The variable conjugate to this holonomy is the holonomy
around the other (contractible) cycle, which is {\em not} fixed
any more to the classical value given by $\Theta = 2\pi$ for
de Sitter space. We must sum over contributions from all possible
values of $\Theta$ in the partition function.
This corresponds to starting with the value of $u$ for the classical solution,
i.e. with $\Theta = 2\pi$
in (\ref{trh}), and then considering all other shifts of $u$ of the form
\begin{eqnarray}
u ~\rightarrow~ u + \alpha \tau
\end{eqnarray}
where $\alpha$ is an arbitrary number. This is implemented by a translation operator
of the form                           
\begin{eqnarray}
T ~=~ \exp \left(\alpha \tau \frac{\partial}{\partial u}\right)
\end{eqnarray}
However, this operator is not gauge invariant. The only gauge-invariant way of implementing these
translations is through Verlinde operators of the form
\begin{eqnarray}
W_{j}~ = ~\sum_{n \in \Lambda_{j}} \exp \left(\frac{-n \pi \bar \tau u}{\tau_{2}} +
\frac{n \tau}{k+2} \frac{\partial}{\partial u} \right)
\end{eqnarray}
where $\Lambda_{j} ~=~ {-j, -j+2,...,j-2, j}$.
This means that all possible shifts in $u$ are not allowed. The
only possible shifts allowed by gauge invariance are:
\begin{eqnarray}
u ~\rightarrow ~u + \frac{n \tau}{k+2}
\label{ushift}
\end{eqnarray}
where $n$ is always an integer taking a maximum value of $k$.
Thus, the only allowed values of $\Theta$ are $2\pi(1 + \frac{n}{k+2})$.
We know that acting on the state with no Wilson lines in the solid torus with the
Verlinde operator $W_{j}$
corresponds
to inserting a Wilson line of spin $j/2$ around the non-contractible cycle.
Thus, taking into account all states with different
shifted values of $u$ as in $(\ref{ushift})$   
means that we have to take into
account all the states  in the boundary corresponding to the insertion
of such Wilson lines. These are the
states $\psi_{j}(u, \tau)$ given in (\ref{c3}).
 
(b) In order to obtain the final partition function, we must also integrate over all
values of the modular parameter $\tau$, i.e. over all inequivalent tori
with the same holonomy around the non-contractible cycle.
The integrand, which is a function of $u$ and $\tau$,
must be the square of the partition function of a
gauged $SU(2)_{k}$ Wess-Zumino model  corresponding to the two $SU(2)$
Chern-Simons theories.
It must be modular invariant -- modular invariance corresponds to
invariance under large diffeomorphisms
of the torus.
The partition function is then of the form
\begin{eqnarray}
Z ~= ~ \int d\mu(\tau, \bar \tau) ~\left|~\sum_{j=0}^{k} ~a_{j}(\tau)~ \psi_{j}
(u, \tau)~\right|^2
\label{pf}
\end{eqnarray}
where $d\mu(\tau, \bar \tau)~ =~ \frac{d\tau d\bar \tau}{\tau_{2}^{2}}$ is the
modular invariant measure, and the integration is over a
fundamental domain in the $\tau$ plane.
Coefficients $a_{j}(\tau)$ must be chosen  such that the
integrand is modular invariant. 
These coefficients are given by
$a_j(\tau)~=~(\psi_j(0,\tau))^*$ so that the partition function is uniquely
fixed to be
\begin{eqnarray}
Z = \int d\mu(\tau, \bar \tau) \left|~\sum_{j=0}^{k}
~(\psi_{j}(0, \tau))^{*} ~\psi_{j}(u, \tau)~\right|^2
\label{bhpf}
\end{eqnarray}
This is an {\em exact} expression for the canonical partition function
in both the cases of BTZ and de Sitter spaces but with appropriate 
identification of holonomies.

To make a comparison with  the semiclassical entropy of
black hole,
we evaluate the expression (\ref{bhpf}) for large  horizon radius $r_{+}$
by the saddle-point method.
Substituting from (\ref{c3}), (\ref{c4}) and (\ref{theta}),
the saddle point of the integrand occurs when $\tau_{2}$ is
proportional to $r_{+}$ and therefore large. But for $\tau_{2}$ large, the character
$\chi_{j}$ is
\begin{eqnarray}
\chi_{j}(\tau, u)~ \sim~ \exp \left[\frac{\pi i \left(\frac{(j+1)^2}{k+2} -
\frac{1}{2}\right)}{2} \tau \right] ~\frac{\sin \pi (j+1)u}{\sin \pi u}
\label{asym}
\end{eqnarray}
We now use in (\ref{bhpf}) the form of the character for 
large $\tau_{2}$  from (\ref{asym}).
In the expression for $u$ in (\ref{udefn}), we replace
$\Theta$ by its classical value $2\pi$. The computation has been done
with positive coupling constant $k$ and at the end,
we must perform an analytic continuation to the Lorentzian black hole, by taking
$G \rightarrow -G$. It can be checked that after the analytic continuation, it is the 
spin $j=0$ in the sum over characters in (\ref{bhpf}) 
that dominates the partition function.               
We obtain the leading behaviour of the partition 
function (\ref{bhpf}) for large $r_{+}$ (and
when $|r_{-}|<< r_{+}$) by first performing the integration over $\tau_{1}$ in this regime.
The $\tau_{2}$ integration is done by the method of steepest descent. The saddle-point is at
$\tau_{2}~=~r_{+}/l$.
Expanding around the saddle-point,
by writing $\tau_{2}~=~r_{+}/l ~+~ x $ and then  integrating over $x$, we obtain
\begin{eqnarray}
Z_{bh} ~=~\frac{l^2}{r_{+}^2} ~\exp \left(\frac{- 2\pi k r_{+}}{l}\right) ~\int dx
~\exp \left[-\frac{\pi k l}{2 r_{+}}~ x^2 \right]
\end{eqnarray}
The integration produces a factor proportional to $\sqrt{r_+}$.
The
partition function for the Lorentzian black hole
of large horizon area $2 \pi r_{+}$ after the analytic continuation $G \rightarrow -G$ is then
\begin{eqnarray}
Z_{Lbh} ~=~ \frac{l^2}{r_{+}^2}~ \sqrt{\frac{8 r_{+} G}{\pi l^2}} ~\exp
\left(\frac{2\pi r_{+}}{4 G}\right)
\label{spbhpf}
\end{eqnarray}
upto a multiplicative constant. The logarithm of this expression
yields the black hole entropy for large horizon length $r_{+}$:
\begin{eqnarray}
S~ = ~\frac{2 \pi r_{+}}{4 G}~ - ~\frac{3}{2} \log \left({\frac{2 \pi r_{+}}
{4G}}\right)~ +~ .~.~.~.
\label{bhentropy}
\end{eqnarray} 
For the case of de Sitter space we compute the partition
function by substituting in the expression (\ref{bhpf}) the values of
$u$ and $\bar u$ with $\Theta = 2\pi$. We work
in the regime where $k$ (and therefore $l$) is large. Also, we must
perform an analytic continuation to get the Lorentzian result - this is
done by taking $G \rightarrow -G$, and $\beta \rightarrow i\beta$.
For the regime when $k$ is large, the leading contribution to the
sum in the integrand comes from $j=0$ as in \cite{btzlog}. The $\tau_{2}$
integral can in fact be done exactly.
We have
\begin{eqnarray}
Z_{dS} = \int_{-1/2}^{1/2} d\tau_{1}~ 4\pi~~ e^{\beta~ k/2l}~~ \frac{1}{f(\tau_{1})}~~ K_{1}(-k/2~~ f(\tau_{1}))
\label{tau1int}
\end{eqnarray}
where $f(\tau_{1}) = \sqrt{ \frac{\beta^2}{l^2} - 4\pi^2 \tau_{1}^{2}}$, and $K_{1}$
is the Bessel function
of imaginary argument.
Using the approximation for the Bessel function with large argument
\begin{eqnarray}
K_{1}(z) = \sqrt{\frac{\pi}{2 z}} e^{-z} [1 + O(\frac{1}{z}) +...]
\label{asyk}
\end{eqnarray}
with the replacement $\beta = 2\pi l$ for de Sitter space, we get, in the large $k$ regime :
 
\begin{eqnarray}
Z_{dS} = 4 \sqrt{\pi}~~ \frac{4G}{2\pi l}~~ e^{2\pi l/4 G}
\label{cpf}
\end{eqnarray}                
Since this is the partition function in the canonical
ensemble, we would have expected an additional term $e^{-i\beta E}$
where $E$ is the energy of
de Sitter space. The notion of energy in asymptotically de Sitter spaces needs to be
defined carefully,
due to the absence of a global timelike Killing vector.
The energy $E$ that emerges in our formalism is defined on the horizon,
and not at asymptotic
infinity, as has been done, for e.g in \cite{vb}.
Our result seems to indicate that that energy
$E$ is zero for de Sitter space.
Such a result coincides with the definition of energy
as given by Abbott and Deser \cite{AD}.                              
The entropy is therefore
\begin{eqnarray}
S = (2\pi l)/4G ~-~\log \frac{2\pi l}{4G} + ........
\label{entropy}
\end{eqnarray}
The leading term is the semi-classical Bekenstein-Hawking entropy
that is proportional to the horizon ``area". The second term is the
leading correction that is logarithmic in area. 

The numerical coefficient of the logarithmic term for BTZ blackhole was
$-3/2$ whereas for the de Sitter case, it is $-1$.
This is somewhat puzzling at first glance. The
black hole entropy was computed in the regime $r_{+} >> l$, where
$r_{+}$ is the black hole horizon radius and $l$ is the $AdS$ radius of
curvature. Then, there was an integral over the modular
parameter similar to (\ref{bhpf}). The saddle-point for $\tau_{2}$,
the imaginary part of the modular parameter occured when $\tau_{2} =
r_{+}/l$. Thus this was the regime when $\tau_{2}$ was large.
An interesting
observation was made in \cite{btzlog} that replacing $r_{+}/l$
in the black hole partition function
by $l/r_{+}$, where now $r_{+} << l$, the $AdS$ gas partition function
was obtained, with the coefficient of the correction being $+3/2$.
This corresponds to a situation where the modular parameter
$\tau_{2} = r_{+}/l$ is small. What happens when $r_{+} \sim l$,
i.e $\tau_{2} \sim 1$? In fact, this is very similar to the de Sitter
case, since the de Sitter horizon radius is exactly $l$! The computation
follows similar lines and leads to similar results. It can in fact
be verified directly from (\ref{bhpf}) that the saddle-point is at
$\tau_{2} = 1$. Here, we see that the coefficient of the logarithmic
correction is $-1$. Thus, the coefficient of the correction seems to depend
on the regime one is looking at. When, as in the above case, there are two
independent length parameters $l$ and $r_{+}$, only for $r_{+} >> l$ do we
get the coefficient $-3/2$

\noindent Summarising our result for BTZ black hole we find:
\begin{eqnarray}
~For~~r_+ >>~l \qquad S~&=&~\frac{2\pi r_+}{4G} ~-~\frac{3}{2}~
\log~( \frac{2\pi r_+}{4G}) ~+~\cdots \nonumber \\
~~~~~r_+ =~l \qquad S~&=&~\frac{2\pi r_+}{4G} ~-~
\log~(\frac{2\pi r_+}{4G})~+~\cdots \nonumber \\
~~~~~r_+ <<~l \qquad S~&=&~\frac{2\pi l^2}{4r_+ G}
~+~\frac{3}{2}~\log(\frac{r_+}{l}) ~+~\cdots
\label{regimetb}
\end{eqnarray}
where the last expression in (\ref{regimetb}) for $r_{+} <<~l$ is the
entropy of the AdS gas.
 
The above results are reminiscent of a duality proposed in \cite{corichi}
between the Euclidean BTZ black hole and Lorentzian de Sitter spaces.

Entropy of de Sitter space can also studied  from an
alternative point of view by using dS/CFT correspondence \cite{strominger}.
In this framework all the information about quantum gravity in the bulk is
expected to be contained in the conformal field theory at
past or future infinity.
The CFT is described by considering all possible metric fluctuations keeping the
asymptotic behaviour to be de Sitter space. It consists of two copies
of Virasaro algebras, each with central charge
$c = 3l/2G$. As shown in \cite{vb}, the
eigenvalues of the Virasoro generators
$L_{0}$ and $\bar L_{0}$ for de Sitter space are
both equal to $l/8G$. Using the Rademacher expansion for modular forms, one can
generalize the Cardy formula for growth of states in a CFT beyond the leading term.
It has been shown \cite{bsen} the sub-leading correction to the entropy
of a BTZ black hole can be determined from this generalisation.
We use these results to find the sub-leading
corrections to the de Sitter entropy from the dS/CFT correspondence.
The entropy obtained from a CFT with a
given the central charge $c$ and eigenvalue of the
Virasoro generator $L_0~=~N$, is given by \cite{bsen}
\begin{eqnarray}
S_1 = S_{0} - 3/2 \log S_{0} + \log c + .......
\label{entr}
\end{eqnarray}
where $S_{0} = 2\pi \sqrt{\frac{c}{6}(N - \frac{c}{24})}$.
This is the contribution from the Virasoro generator $L_{0}$. There is a similar
contribution $S_2$ associated with the Virasoro generator
$\bar L_{0}$, given by replacing $N$
in the above by $\bar N$, the eigenvalue of $\bar L_{0}$.
                                                          
Substituting
$c = 3l/2G$ and $N = \bar N = l/8G$ in the above, we see that
\begin{eqnarray}
S = S_1 + S_2 = 2\pi l/4G - \log \frac{2\pi l}{4G} + ....
\label{cftentr}
\end{eqnarray}
with the same coefficient $-1$ for the logarithmic correction as that obtained
from the gravity partition function (\ref{cpf}) in (\ref{entropy}). 

Thus, the quantum gravity calculation of de Sitter entropy and the
entropy computation from the asymptotic CFT agree even in the sub-leading
correction to the Bekenstein-Hawking term.

\nonumsection{References}

\end{document}